# Manipulating the hydrogen-induced insulator-metal transition through artificial microstructure engineering

*Xuanchi Zhou [1, 2] † \*, Xiaohui Yao [1] †, Wentian Lu [1, 2] †, Jinjian Guo [1], Jiahui Ji [1], Lili Lang [3], Guowei Zhou [1, 2] \*, Chunwei Yao [1], Xiaomei Qiao [1], Huihui Ji [1, 2], Zhe Yuan [4], Xiaohong Xu [1, 2] \**

[1] *Key Laboratory of Magnetic Molecules and Magnetic Information Materials of Ministry of Education & School of Chemistry and Materials Science, Shanxi Normal University, Taiyuan, 030031, China*
[2] *Research Institute of Materials Science, Shanxi Key Laboratory of Advanced Magnetic Materials and Devices, Shanxi Normal University, Taiyuan 030031, China*
[3] *National Key Laboratory of Materials for Integrated Circuits, Shanghai Institute of Microsystem and Information Technology, Chinese Academy of Sciences, Shanghai 200050, China*
[4] *Interdisciplinary Center for Theoretical Physics and Information Sciences, Institute of Nanoelectronic Devices and Quantum Computing, Fudan University, Shanghai 200433, China*

\*Authors to whom correspondence should be addressed: *xuanchizhou@sxnu.edu.cn (X. Zhou)*, *zhougw@sxnu.edu.cn (G. Zhou)*, and *xuxh@sxnu.edu.cn (X. Xu)*.

*† X. Zhou, X. Yao, and W. Lu contributed equally to this work.*

**Abstract**

Hydrogen-associated filling-controlled Mottronics within electron-correlated system provides a groundbreaking paradigm to explore exotic physical functionality and phenomena. Dynamically controlling hydrogen-induced phase transitions through external fields offers a promising route for designing protonic devices in multidisciplinary fields, but faces high-speed bottlenecks owing to slow bulk diffusion of hydrogens. Here, we present a promising pathway to kinetically expedite hydrogen-related Mott transition in correlated $VO_2$ system by taking advantage of artificial microstructure design. Typically, inclined domain boundary configuration and $c_R$-faceted preferential orientation simultaneously realized in $VO_2/Al_2O_3$ ($1\bar{1}02$) heterostructure significantly lower the diffusion barrier via creating an unobstructed conduit for hydrogen diffusion. As a result, the achievable switching speed through hydrogenation outperforms that of counterpart grown on widely-reported *c*-plane $Al_2O_3$ substrate by 2-3 times, with resistive switching concurrently improved by an order of magnitude. Of particular interest, an anomalous uphill hydrogen diffusion observed for $VO_2$ with a highway for hydrogen diffusion fundamentally deviates from basic Fick's law, unveiling a deterministic role of hydrogen spatial distribution in tailoring electronic state evolution. The present work not only provides a versatile strategy for manipulating ionic evolution, endowing with great potential in designing high-speed protonic devices, but also deepens the understanding of hydrogen-induced Mott transitions in electron-correlated system.

**Key words**: Correlated oxides, Ionic evolution, Electronic phase transition, Microstructure engineering, Hydrogen diffusion;

# 1. Introduction

Ionic control of correlated oxides opens a pioneering paradigm to dynamically manipulate multiple physical functionalities and discover exotic physical phenomena, enabling multidisciplinary applications in energy conversions, artificial intelligence, superconductivity, bio-sensing, and correlated electronics.[1-8] Among the existing ionic species, hydrogen ions or protons (e.g., $H^+$), characterized by the smallest ionic radius and ultrahigh mobility, can be exploited to trigger topotactic phase modulations within correlated oxide system in a more reversible and controllable fashion.[9] Incorporating hydrogens into the lattice of correlated oxides can bring in an additional ion degree of freedom to promote the complex interplay of charge, lattice, spin and orbital degrees of freedom, serving as a promising platform to probe ion-electron-lattice coupling.[4, 10-13] As a representative case, hydrogenation provides a powerful pathway to induce Mott phase modulations in an extensive collection of $d$-orbital correlated oxide system, for example, $VO_2$, $SrRuO_3$ and $ReNiO_3$, leading to a rich spectrum of magnetoelectric states.[14-23] The protonic control over the electronic states of correlated oxides is closely associated with the hydrogen-associated band-filling process, during which doped electron carriers adjust the band structure to upset the stability of correlated electronic ground state.[24] More importantly, manipulating the hydrogen-mediated phase transitions using external fields provides a fertile ground to develop proton-based device applications.[25-26] An archetypical example is the protonic transistor device, in which the voltage-actuated proton evolution can reversibly adjust the channel conductance, triggering multi-state resistive switching.[27-28]

Nevertheless, the relatively slow hydrogenation kinetics resulting from the bulk diffusion of hydrogens poses a significant challenge to designing high-speed electronic devices using proton evolution. Conventionally, the hydrogen diffusion within the lattice of correlated oxides is driven by the hydrogen concentration gradient, following Fick's laws of diffusion, the kinetics of which is progressively depressed with the diffusion length away from the hydrogen source.[9] One feasible pathway for promoting the proton evolution is to artificially engineer the material microstructure that establishes a highway for hydrogen diffusion via lowering the diffusion barrier. Typically, intercalated hydrogens, acting as interstitial point defects, readily interact with extended planer defects in correlated oxides, and therefore an inclined boundary configuration accelerates the hydrogen diffusion, in comparison with the horizontally aligned one.[24, 29-31] In addition, hydrogens are prone to diffuse along an empty channel within the lattice of correlated oxides, rendering the crystal facet anisotropy in hydrogen-related electronic phase modulations.[32] Delicately designing the material microstructure is poised to create unblocked conduits for hydrogen diffusion, allowing for an enhanced hydrogen diffusivity. From a microscopic perspective, the ability to control the hydrogen spatial distribution offers a powerful tool for precisely tailoring hydrogen-related electronic state evolution in an electron-correlated system. Establishing the relationship between microscopic hydrogen distribution and macroscopic electronic phase modulation not only deepens the understanding of how the protons behave in phase transition but also drives forward protonic device applications.

Here, we identified VO$_2$, an electron-correlated system exceptionally responsive to external stimuli, as an ideal platform for adjusting proton evolution through delicately designing material microstructure. Conventionally, VO$_2$ undergoes an abrupt insulator-metal transition that is driven by a critical temperature of 341 K, accompanied by a symmetry-lowering structural transformation from low-temperature monoclinic phase to high-temperature rutile phase.[33-34] Beyond thermally-driven phase transition, hydrogenation triggers sequential insulator-metal-highly insulator Mott phase transition in VO$_2$ system, endowing with an isothermal resistive switching function.[14, 24] In this work, the artificial design of material microstructure is identified as an effective pathway to kinetically expedite proton evolution via establishing an unobstructed freeway for hydrogen diffusion. In particular, an abnormal uphill hydrogen distribution, opposite to the hydrogen concentration gradient, was observed for VO$_2$ with an unobstructed conduit, indicating a close relationship between microscopic hydrogen distribution and macroscopic phase transition. This work highlights the robust capability of manipulation on the proton evolution of VO$_2$ by taking advantage of the microstructure engineering, offering an promising pathway for designing high-speed protonic devices.

## 2. Results and discussion
### Hydrogen-associated structural evolution in metastable VO$_2$ (B)

Hydrogen-associated electron-doping Mottronics triggers tri-state orbital reconfiguration of correlated VO$_2$ from correlated electronic ground state ($t_{2g}^1 e_g^0$) to either electron-itinerant state ($t_{2g}^{1+\Delta} e_g^0$) or electron-localized state ($t_{2g}^2 e_g^0$), transcending temperature-controlled phase transition (Figure 1a).[14, 24] Introducing the electron carriers to partially occupy the low-energy $d_{//}^*$ orbital through proton evolution triggers the metallization of VO$_2$. By strong contrast, with excessive hydrogenation, the complete filling in the $d_{//}^*$ orbital of VO$_2$ instead opens a wider band gap between $d_{//}^*$ and $\pi^*$ orbitals, leading to the electron localization via on-site Coulomb repulsions. This breakthrough enables the possibility in designing electronic device applications using the proton evolution, but realizing high-speed operation remains a technical challenge due to the inherently slow bulk diffusion kinetics. To facilitate proton evolution, it is highly desirable to delicately design the material microstructure for building up an unobstructed conduit for hydrogen diffusion, which can significantly reduce the hydrogen diffusion barrier (Figure 1b). Notably, the hydrogen-related electronic phase modulations in correlated VO$_2$ system are closely correlated with the incorporated hydrogen concentration that reconfigures the $d$-orbital electronic band structure. Therefore, the artificial design in material microstructure is expected to not only expedite hydrogenation kinetics but also tailor resultant electronic phase transitions, hinting at great potentials for surmounting the primary challenge in developing advanced protonic devices.

To address the above central concept, the differently oriented rutile $TiO_2$ and hexagonal $Al_2O_3$ are intentionally selected as epitaxial template for designing the domain boundary configuration and crystal orientation of rutile $VO_2$ films. Noting similar *a*-axis lattice constant and identical rutile structure between $VO_2$ and $TiO_2$, the rutile-on-rutile coherent epitaxial growth of $VO_2$ film deposited on the $TiO_2$ (100) substrate is expected to engender the horizontally-aligned domain boundary in $VO_2$ film that tightly conforms to the substrate orientation. Such the coherent epitaxy in $VO_2/TiO_2$ heterostructure starkly differs from the situation of rutile $VO_2$ films grown on hexagonal $Al_2O_3$ substrate, in which case the symmetry mismatch is expected to induce potential lattice rotation and twinning tilt of $VO_2$, incurring a domain-matching epitaxial growth. This understanding is further identified by the high-resolution transmission electron microscopy (HRTEM) in Figure 1c, where the vertically-aligned domain boundaries are visualized in as-deposited $VO_2/Al_2O_3$ (0001) heterostructure (Figure 1d). The symmetry mismatch between hexagonal $Al_2O_3$ and rutile $VO_2$, coupled with the $\beta$ angular deviation of $VO_2$ (e.g., 122.6 °) from the $Al_2O_3$ (e.g., 120 °), leads to three equivalent sets of twin variants in $VO_2$ film (Figure 1e). It is found that the atomic arrangement in set 1 resembles the (001) plane of rutile $VO_2$, while the zigzag V-V chain in set 2 (e.g., $[011]_R$ domain) related to the dimerization of $VO_2$ results in a lowering crystal symmetry, in comparison with the $[01\bar{1}]_R$ domain of $VO_2$ in set 3 (Figure S1).[35] By contrast, a 45 °-tilted twin boundary relative to the heterointerface was observed for as-deposited $VO_2/Al_2O_3$ (1$\bar{1}$02) heterostructure in Figures 1f-1h. This phenomenon is ascribed to two equivalent (200)-faceted and ($\bar{2}$11)-faceted domains existed in the $VO_2$ film grown on the *r*-plane $Al_2O_3$ substrate, in which the angle between (200) and ($\bar{2}$11) planes as estimated to be 44.8 ° aligns well with the observed twinning tilt of ~ 45 °. In addition, the relatively smooth surface roughness for as-deposited $VO_2$ films is characterized by using the atomic force microscope in Figure S2. Consequently, the artificial design in the domain boundary of $VO_2$ with diverse orientations relative to the heterointerface, ranging from 0 ° to 90 °, is realized through elaborately selecting as-used epitaxial templates. Such the precise control over the domain boundary configuration is expected to effectively tailor the proton evolution, in which inclined domain boundary configuration (e.g., 45 ° or 90 °) acts as a freeway for promoting hydrogen diffusion via ionic interactions between hydrogens and extended planar defects.

On this basis, a platinum (Pt)-assisted hydrogen spillover strategy was exploited for achieving the hydrogen ions intercalation into the lattice of $VO_2$. The noble metal Pt as catalyst can substantially reduce the energy barrier for dissociating the hydrogen molecules derived from the $H_2/Ar$ forming gas into the protons and electrons at the triple phase boundary, triggering a catalytic reaction associated with $H_2$ (g)→$H^+$+$e^-$.[10, 36] To compare the hydrogenation kinetics of $VO_2$ under microstructure design, a fairly mild hydrogenation condition (e.g., 70 ° C, 30 min) was herein employed, in comparison with the previous reports,[9, 17, 24] which simultaneously avoids the formation of oxygen defects. From the perspective of structural evolution, performing the hydrogenation results in the leftward shift of characteristic diffraction peak of $VO_2$ in their X-ray diffraction (XRD) patterns, regardless of selected epitaxial templates (Figure 2a). Such the hydrogen-triggered structural evolution of VO$_2$ reveals an *out-of-plane* lattice expansion, primarily ascribed to the O-H interactions. It is worthy to note that the grown VO$_2$ films on both the (0001)-oriented Al$_2$O$_3$ and (100)-oriented TiO$_2$ substrates are (100)$_R$-faceted crystal orientation, while in contrast the *r*-plane Al$_2$O$_3$ template can induce the preferential growth of VO$_2$ toward (100)$_M$ orientation, equivalent to the $c_R$ orientation. It is widely reported that a lower energy barrier for hydrogen diffusion is achieved along the [001]$_R$ channel of VO$_2$ compared to the [100]$_R$ channel, owing to an empty channel along the $c_R$ direction as formed by the chains of edge-sharing VO$_6$ octahedra.[37-38] Therefore, the *r*-plane Al$_2$O$_3$ template, which can simultaneously induce a $c_R$-faceted crystal orientation and tilted domain boundary configuration within the lattice of VO$_2$ film, dramatically reduces the hydrogen diffusion barrier.

The effective regulation in hydrogen-triggered insulator-metal transition through microstructure design is confirmed by comparing the temperature dependence of material resistivity in Figure 2b. With a mild hydrogenation at 70 °C for 30 min, the VO$_2$/TiO$_2$ (100) heterostructure still retains a pronounced thermally-driven phase transition, akin to pristine state. By contrast, the electronic phase transition of VO$_2$/Al$_2$O$_3$ (0001) bilayer is dramatically depressed upon hydrogenation, resulting in a transition sharpness ($\rho_{Insul.}/\rho_{Metal.}$) below $10^1$. Of particular note is the complete metallization for the VO$_2$ film deposited on *r*-plane Al$_2$O$_3$ substrate, identifying an accelerated hydrogenation kinetics through microstructure design. An inclined boundary configuration and preferential $c_R$-faceted orientation as simultaneously realized in the VO$_2$/Al$_2$O$_3$ (1$\bar{1}$02) heterostructure facilitate the proton evolution via lowering the diffusion barrier of hydrogens, thereby enhancing resultant electronic phase modulation (Figure S3a). This understanding is further confirmed by comparing the hydrogen-triggered resistive regulation (e.g., $R_0/R_H$) of VO$_2$ under microstructure engineering, in which a more pronounced $R_0/R_H$ is realized in VO$_2$/Al$_2$O$_3$ (1$\bar{1}$02) bilayer as shown in the inset of Figure 2b. The above regulation on the proton evolution through microstructure design is reproducible (Figures S3b-S3c) and analogously observed in VO$_2$ hydrogenated at 70 °C for 10 min (Figure S4). Further consistency is verified by respective dehydrogenation process in Figure 2c via exposing hydrogenated VO$_2$ to the air, during which the intercalated hydrogens within VO$_2$ tend to be dragged out owing to a low crystal trapping potential and ultrahigh chemical diffusivity. Similarly, corresponding dehydrogenation process is effectively adjusted by using microstructure engineering, wherein the dehydrogenation kinetics of VO$_2$/Al$_2$O$_3$ (1$\bar{1}$02) bilayer is expedited, evidenced by a reduced onset time for triggering resistive recovery and an elevated magnitude of $R_H/R_0$. It is also worthy to note that hydrogen-triggered phase transition and structural evolution within VO$_2$ system are highly reversible when annealed in the air at 70 ºC for 30 min, in alignment with the prior report (Figures S5-S6).[2]

In addition, hydrogen-associated structural transformation is further determinated using the Raman spectra in Figure 2d, where the Raman peaks at 194, 223, and 612 cm$^{-1}$ can be taken as a direct indicator of structural transformation in VO$_2$ system. The Raman peaks located at 194 cm$^{-1}$ ($\omega$1) and 223 cm$^{-1}$ ($\omega$2) correspond to the interaction of V-V dimers, while the Raman peak at 612 cm$^{-1}$ ($\omega$3) reflects the difference in the V-O bond length.[2] Such the characteristic Raman peaks are almost suppressed for VO$_2$/Al$_2$O$_3$ (1$\bar{1}$02) heterostructure through hydrogenation, revealing the conversion to a rutile-like crystal structure, accompanied by the suppression of V-V dimers. This observation differs from the case of VO$_2$/Al$_2$O$_3$ (0001) bilayer, in which situation the above Raman peaks (e.g., $\omega$ 1 and $\omega$ 2 peaks) still remain detectable upon hydrogenation, signifying remnant monoclinic phase that cannot transit to rutile phase. Without loss of generality, a non-catalytic hydrogenation strategy leveraging the electron-proton co-doping was exploited to generalize the artificial microstructure design strategy for adjusting the proton evolution.[39] In this scenario, the Fermi level difference between the low work-function metal (e.g., Al) and VO$_2$ induces a spontaneous electron transfer from Al particle to the VO$_2$ film, which attracts the protons from the acid solution to be intercalated into the lattice of VO$_2$, realizing the hydrogenation (Figure S7).[39] Hydrogen-associated structural evolution of VO$_2$ using acid solution strategy is demonstrated by XRD spectra (Figure S8). More importantly, the onset time for triggering the resistive reduction in VO$_2$/Al$_2$O$_3$ (1$\bar{1}$02) heterostructure through hydrogenation is significantly lower than that for VO$_2$/Al$_2$O$_3$ (0001) hybrid by 2-3 times, indicative of an accelerated kinetics of proton evolution (Figure 2e). Meanwhile, the hydrogen-triggered resistive switching (e.g., $R_0/R_H$) achievable in VO$_2$/Al$_2$O$_3$ (1$\bar{1}$02) hybrid largely exceeds the one for VO$_2$ deposited on the $c$-plane Al$_2$O$_3$ by almost one order of magnitude. Establishing an unobstructed conduit through artificially designing the material microstructure kinetically promotes the hydrogen diffusion and associated electronic phase modulations of VO$_2$.

To clarify the variations in chemical environment of VO$_2$ under hydrogenation, X-ray photoelectron spectra (XPS) analysis was performed, as the V $2p$ core-level spectrum shown in Figure 3a. It is found that the valence state of vanadium is reduced from V$^{4+}$ to V$^{(4-\delta)+}$ upon hydrogenation, particularly for the VO$_2$/Al$_2$O$_3$ (1$\bar{1}$02) hybrid. In addition, performing the hydrogenation elevates the relative intensity of the O-H interaction (e.g., ~532 eV) for VO$_2$ with respect to the V-O interaction (e.g., ~530 eV) (Figure 3b).[16] The O-H interaction as formed in hydrogenated VO$_2$ indicates that the intercalated hydrogens acting as interstitial defects readily bond with the lattice oxygen, forming the O-H weak bonds, in accordance with previous XRD results. Furthermore, a more pronounced elevation in the O-H interaction was observed for hydrogenated VO$_2$/Al$_2$O$_3$ (1$\bar{1}$02) heterostructure in comparison with the VO$_2$/Al$_2$O$_3$ (0001) (Figure S9), which further demonstrates an overwhelming advantage of a freeway in promoting proton evolution created by a 45 º-tilted twin boundary and preferential $c_R$ orientation.

In order to provide more direct evidences for hydrogen spatial distribution within the lattice of VO$_2$, time-of-flight secondary ion mass spectrometry (ToF-SIMS) analysis was performed to semiquantitatively characterize the depth profiles of hydrogen concentration. ToF-SIMS analysis as a destructive technique mainly utilized an ion beam (e.g., $Ar_n^+$ and $Cs^+$) to sputter through hydrogenated $VO_2$ film, ejecting the molecular ions that can be further analyzed using a mass spectrometer (Figure 3c).[40] In addition, the change in the intensity of molecular species over sputtering time provides a depth-resolved understanding of incorporated hydrogen concentration, which identifies microscopic hydrogen distribution within the lattice of $VO_2$. In $VO_2$ films deposited on *c*-plane and *r*-plane $Al_2O_3$ substrates, the presence of hydrogen is evident over the film region via three-dimensional ToF-SIMS element maps (Figures 3c and S10), contrasting with $VO_2$/$TiO_2$ (100) bilayer where no significant H signal discrepancy exists between film and substrate regions (Figure S11).

The kinetic acceleration in hydrogen diffusion through the artificial design of material microstructure is clearly demonstrated by comparing the depth profile of hydrogen concentration, as performed using ToF-SIMS analysis in Figure 3d. With a fairly mild hydrogenation (e.g., at 70 ºC for 30 min), a significantly higher hydrogen intensity is observed for the grown $VO_2$ film in comparison with the *c*-plane $Al_2O_3$ substrate, unveiling an effective hydrogen incorporation, which differs from $VO_2$/$TiO_2$ (100) hybrid with a similar H intensity over the entire heterostructure region. The vertical domain boundary configuration in $VO_2$/$Al_2O_3$ (0001) bilayer promotes the hydrogen diffusion, compared with the $VO_2$/$TiO_2$ (100) with horizontally-aligned domain boundary, in agreement with previous study (Figures S12-S13).[41] Surprisingly, an uphill hydrogen distribution is observed for $VO_2$/$Al_2O_3$ ($1\bar{1}02$) bilayer, where the progressive increase in hydrogen concentration with diffusion length fundamentally deviates from basic Fick's law of diffusion (Figure S14). This phenomenon stands stark contrast with conventional understanding of hydrogen diffusion, which is driven by the hydrogen concentration gradient as described by the Fick's law of diffusion. Such the anomalous uphill hydrogen diffusion realized in the $VO_2$/$Al_2O_3$ ($1\bar{1}02$) bilayer is associated with the pre-existing unobstructed freeway, in favor of hydrogen diffusion. The averaged hydrogen concentration achievable in $VO_2$/$Al_2O_3$ ($1\bar{1}02$) bilayer also exceeds the one for widely-reported $VO_2$/$Al_2O_3$ (0001) and $VO_2$/$TiO_2$ (100) heterostructures,[24, 27, 39, 41] indicating a kinetically favorable proton evolution that leads to a faster and more abrupt electronic phase modulation (Figure 3e). This result further demonstrates a close correlation between hydrogen spatial distribution and macroscopic phase modulations. In terms of protonic device application, such the uphill hydrogen distribution realized in $VO_2$ through microstructure design is also poised to mitigate the need for significant energy input typically associated with hydrogen supply.

To probe the physical origin driving hydrogen-related electronic phase modulations, the electronic band structure of $VO_2$ is characterized using soft X-ray absorption spectroscopy (sXAS) technique and density functional theory (DFT) calculations in Figure 4. The V-*L* edge spectrum associated with the V $2p \rightarrow 3d$ transition is widely recognized to effectively reflect the variations in the vanadium valence state.[39] Performing the hydrogenation results in the leftward shift of both V-$L_{III}$ and V-$L_{II}$ peaks, unraveling the reduction in the valence state of vanadium (Figure 4a), while such the hydrogen-triggered red shift is more pronounced for VO$_2$/Al$_2$O$_3$ (1 $\bar{1}$ 02) heterostructure, reminiscent of that found in respective XPS spectra. Noting empty O-2$p$ states and the hybridization between V-3$d$ and O-2$p$ orbitals, the unoccupied density of states in the conduction band of VO$_2$ can be reflected by the O 1$s$ spectrum, in which the relative variation in the first (second) peak intensity qualitatively represents the electron occupation in the $t_{2g}$ ($e_g$) band of VO$_2$.[42] Upon hydrogenation, the reduction in the spectral weight of the first peak in O 1$s$ core-level spectrum with respect to the second peak unveils the band filling in the low-energy $t_{2g}$ band of VO$_2$, which is formed by $d_{xz}$, $d_{yz}$ and $d_{x^2-y^2}$ orbitals (Figure 4b). Compared with the VO$_2$/Al$_2$O$_3$ (0001) hybrid, a more abrupt reduction in the ratio of $t_{2g}$/$e_g$ peak observed for VO$_2$/Al$_2$O$_3$ (1 $\bar{1}$ 02) through hydrogenation is ascribed to the kinetically accelerated proton evolution that introduces more extensive electron doping in the $t_{2g}$ orbital of VO$_2$. Analogous variations in the sXAS spectra are realized in VO$_2$ film as hydrogenated at 70 ºC for 10 min (Figure S15). On the basis of electronic band structure, the doped electron carriers prefer to occupy the low-energy $d_{//}^*$ orbital of VO$_2$ to reconfigure the band structure, triggering a collective metallization.

The underneath mechanism governing the hydrogen-induced insulator-metal transition within VO$_2$ is more clearly clarified via analyzing the electronic density of states (DOS) using DFT calculations (Figures 4c and S16). It is found that a band gap of ~ 0.8 eV is observed for pristine VO$_2$, in accordance with its correlated electronic ground state (Figure 4d). However, introducing hydrogens into the lattice of VO$_2$ (e.g., H$_4$V$_8$O$_{16}$ with 0.5 H per u.c.) renders the metallization of VO$_2$, evidenced by the finite DOS near the $E_F$ (Figure 4e). Furthermore, the electron carriers provided by hydrogenation are prone to occupy the V-3$d$ orbital, according to DFT calculations. By virtue of DFT calculations and sXAS analysis, the physical origin underlying hydrogen-induced electronic phase transition is identified as band-filling-mediated orbital reconfiguration. Moreover, utilizing the ultraviolet photoelectron spectroscopy (UPS) technique, the work function of VO$_2$ is demonstrated to be reduced through hydrogenation, e.g., from 5.12-5.85 eV to 4.69-4.74 eV, owing to hydrogen-associated electron doping (Figure S17).

## 3. Conclusion

In summary, we identified a feasible strategy to kinetically accelerate the proton evolution in correlated VO$_2$ system through establishing an unobstructed freeway for promoting hydrogen diffusion using artificial microstructure design. The domain-matching epitaxy growth of VO$_2$ film on Al$_2$O$_3$ substrate resulting from the symmetry mismatch, along with the rutile-on-rutile coherent growth in VO$_2$/TiO$_2$ bilayer, serves as a promising platform for delicately designing the VO$_2$ microstructure. By leveraging a 45 º-tilted boundary configuration and $c_R$-faceted crystal orientation, simultaneously induced by *r*-plane $Al_2O_3$ template, the diffusion barrier of hydrogens is extensively reduced, leading to an unusual uphill hydrogen diffusion behavior that violates the Fick's law of diffusion. As a result, hydrogen-related electronic phase modulation achievable in $VO_2$/$Al_2O_3$ ($1\bar{1}02$) hybrid is significantly enhanced compared to the one grown on the widely-reported *c*-plane $Al_2O_3$ substrate, with resistive switching improved by an order of magnitude and switching speed expedited by 2-3 times. Our findings provide compelling evidence for the close correlation between hydrogen spatial distribution and macroscopic filling-controlled Mott phase transitions, suggesting a promising pathway for precisely tailoring hydrogenated electronic state. In addition, we showcase the robust capability of accelerating ionic evolution in correlated oxide system by virtue of creating a freeway for hydrogen diffusion through microstructure design, offering a proof-of-principle for designing high-speed protonic devices.

## 4. Experimental Section

*Fabrication of the grown VO$_2$ heterostructures:* The VO$_2$ films were deposited on the single crystalline (100)-oriented TiO$_2$ and (0001) or (1$\bar{1}$02)-oriented Al$_2$O$_3$ substrates using the laser molecular beam epitaxy (LMBE) technique. The deposition temperature, the oxygen pressure, the target-substrate distance and the laser fluence were optimized as 500 ºC, 1.5 Pa, 45 mm and 1.0 J cm$^2$, respectively. Afterwards, the as-deposited VO$_2$ films were naturally cooled down to the room temperature under identical oxygen partial pressure. Before the hydrogenation, the 20 nm-thick platinum dots were sputtered into the surface of the grown VO$_2$ films via exploiting the magnetron sputtering technique. Finally, based on the hydrogen spillover strategy, as-made Pt/VO$_2$ heterostructures were annealed in a 5 % H$_2$/Ar forming gas for effectively realizing hydrogenation. Additionally, a metal-assisted acid solution strategy was also employed to hydrogenate the VO$_2$ films, in which a 1×1×1 mm$^3$ Al particle was placed onto the surface of VO$_2$ films immersed in a dilute sulfuric acid solution.

*Material characterizations:* The crystal structures of the grown VO$_2$ films were probed via using the X-ray diffraction (XRD) (Rigaku, Ultima IV) and Raman spectra (HORIBA, HR Evolution). High-resolution transmission electron microscopy (HRTEM) (JEOL, JEM F200; JEM, ARM300F) was also performed to characterize the crystal structure of VO$_2$ films that were first fabricated via using the focused ion beam (FIB) (FEI, G4 UX). The variations in the chemical environment of hydrogenated VO$_2$ were assessed by using the X-ray photoelectron spectroscopy (XPS) technique (Thermo, K-Alpha X). The electronic structure of VO$_2$ films was further explored through soft X-ray absorption spectroscopy (sXAS) analysis, as conducted at the Shanghai Synchrotron Radiation Facility (SSRF) on beamline BL08U1A. Temperature dependent resistivity of the deposited VO$_2$ films were measured by using a commercial physical property measurement system (PPMS) (Quantum design), meanwhile the room-temperature resistance were determined by using a Keithley 2400 system. The depth profile of hydrogen is examined by using the time-of-flight secondary ion mass spectrometry (ToF-SIMS) technique (ION-TOF GmbH, TOF.SIMS 5).

*First-principles calculations:* First-principles calculations were conducted using the projector augmented wave (PAW) method within the QUANTUM ESPRESSO framework. To model the hydrogenated VO$_2$ (H$_4$V$_8$O$_{16}$), a 2×1×1 supercell was expanded along the *a*-axis, incorporating two hydrogen atoms per unit cell. The computational setup included a plane-wave kinetic energy cutoff of 90 Ry and a 5×11×9 Monkhorst-Pack k-point mesh for sampling the Brillouin zone. Electron correlation effects in the V-3*d* orbitals were accounted for via the GGA+U method,[43] with a Hubbard *U* value of 3.32 eV applied to the V-3*d* states owing to the strong electron correlations. Structural optimization was achieved by relaxing atomic positions until the inter-atomic forces were reduced below 10$^{-3}$ Ry/Bohr. Self-consistent field calculations were performed with an energy convergence criterion of $10^{-12}$ Ry to ensure high numerical accuracy. Finally, the hydrogen-associated electronic band structure was computed based on the optimized geometry, facilitating a detailed investigation into the changes in electronic states induced by hydrogenation.

**Acknowledgements**

This work was supported by the National Natural Science Foundation of China (Nos. 52401240, U24A6002, 52471203, 12404139 and 12174237), Fundamental Research Program of Shanxi Province (No. 202403021212123), Scientific and Technologial Innovation Programs of Higher Education Institutions in Shanxi (No. 2024L145), and the Open Project of Tianjin Key Laboratory of Optoelectronic Detection Technology and System (No. 2024LODTS102). The authors acknowledge the beam line BL08U1A at the Shanghai Synchrotron Radiation Facility (SSRF) (https://cstr.cn/31124.02.SSRF.BL08U1A) for the assistance of sXAS measurement.

**Conflict of Interest**

The authors declare no conflict of interest.

**Figures and captions**

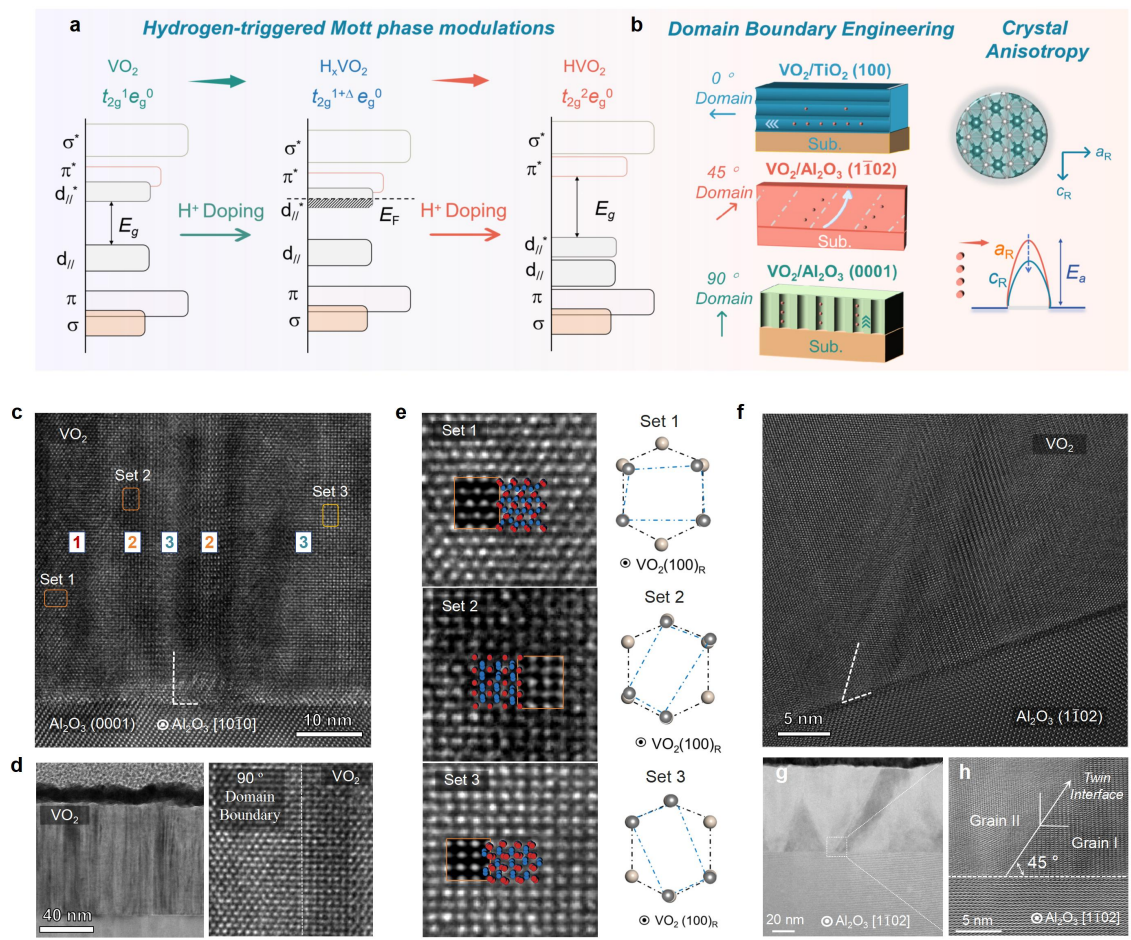

**Figure 1. Artificial design in the material microstructure. a**, Schematic illustration of the hydrogen-triggered Mott phase modulations within VO$_2$ system. **b**, Schematic of artificially engineering the domain boundary and crystal orientation of VO$_2$. **c**, High-resolution transmission electron microscopy (HRTEM) images for as-deposited VO$_2$/Al$_2$O$_3$ (0001) heterostructure. **d**, Visualization of the vertically-aligned domain boundary in VO$_2$ deposited on the *c*-plane Al$_2$O$_3$ substrate. **e**, HRTEM images for three sets of equivalent atomic arrangements of VO$_2$ grown on the *c*-plane Al$_2$O$_3$ substrate. **f**, HRTEM images for as-deposited VO$_2$/Al$_2$O$_3$ (1$\bar{1}$02) heterostructure. **g**, Low-magnification and **h**, high-magnification images for visualizing the 45 º-tilted domain boundary in VO$_2$/Al$_2$O$_3$ (1$\bar{1}$02) bilayer using the HRTEM analysis.

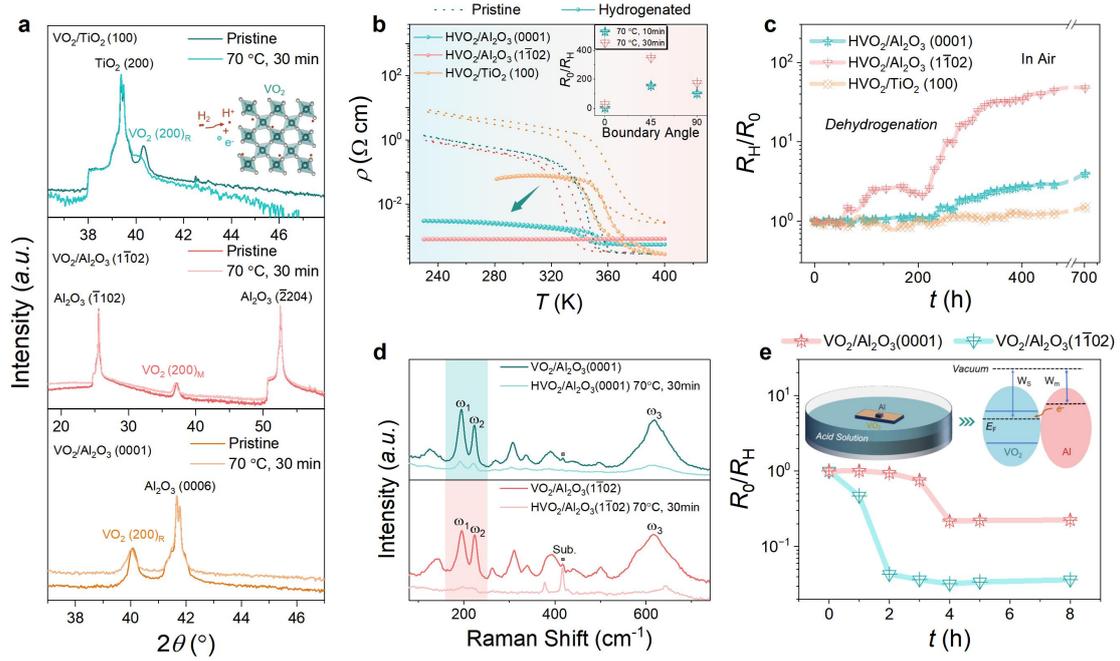

**Figure 2. Hydrogen-related structural evolution and electronic phase transitions.**
**a**, X-ray diffraction (XRD) patterns compared for $VO_2$ films under microstructure engineering through hydrogenation. **b**, Temperature dependence of material resistivity ($\rho$-$T$) as measured for $VO_2$ under microstructure engineering as hydrogenated at 70 ºC for 30 min, while hydrogen-induced variation in the material resistivity ($R_0/R_H$) is shown in the inset. **c**, Dehydrogenation process as compared for $VO_2$ under microstructure engineering via exposing to the air. **d**, Raman spectra as compared for $VO_2$ deposited on the (0001) and (1$\bar{1}$02)-oriented $Al_2O_3$ substrates before and after hydrogenation. **e**, Temporal evolution of the magnitude of $R_0/R_H$ compared for $VO_2$ films deposited on the $c$-plane and $r$-plane $Al_2O_3$ substrates, as hydrogenated using acid solution strategy.

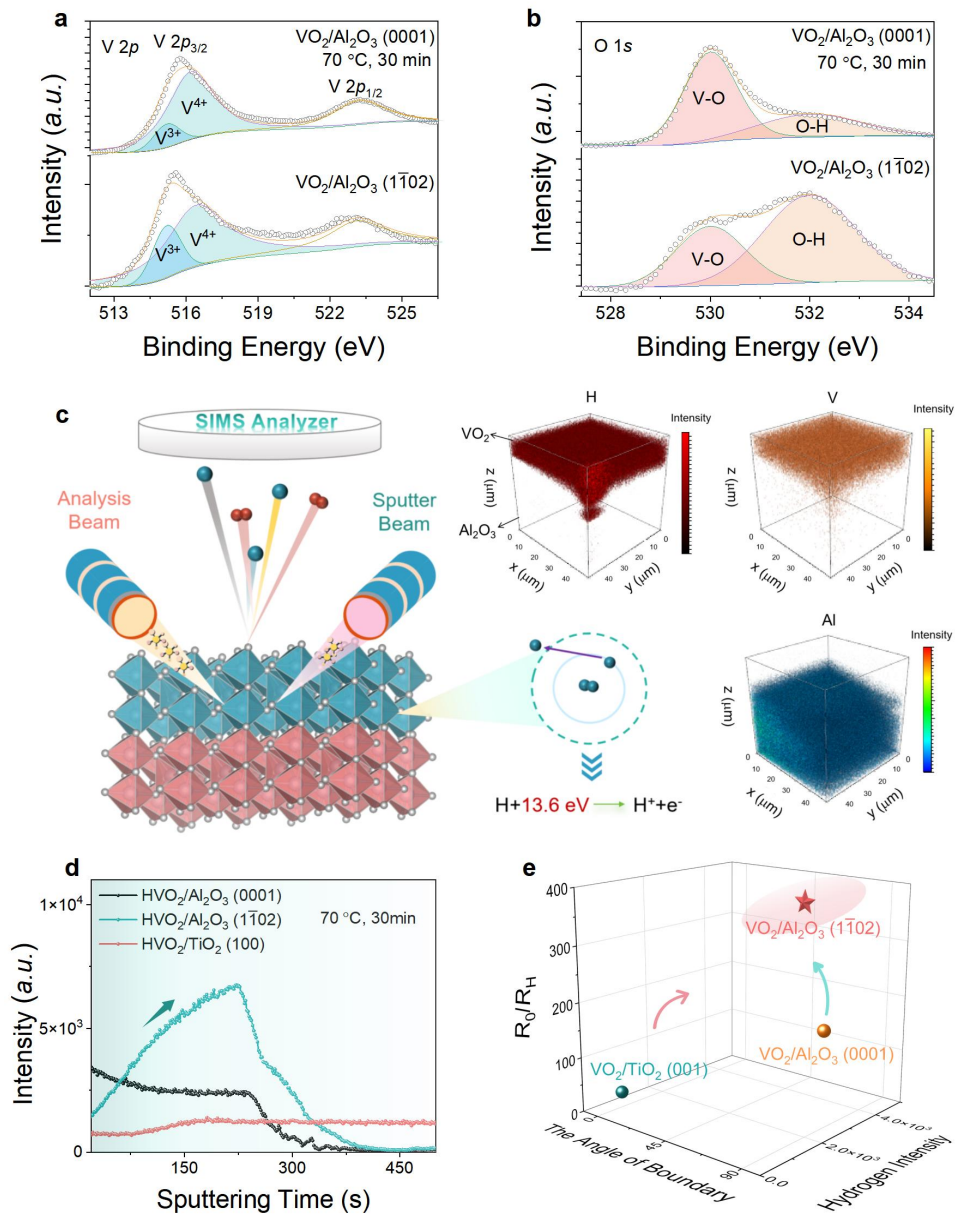

**Figure 3. Hydrogenation kinetics and chemical environment for VO$_2$. a-b**, X-ray photoelectron spectra (XPS) spectra for the core levels of **a**, vanadium and **b**, oxygen of hydrogenated VO$_2$ under microstructure engineering. **c**, Schematic of the working principle of time-of-flight secondary ion mass spectrometry (ToF-SIMS) and three-dimensional ToF-SIMS element maps for hydrogenated VO$_2$/Al$_2$O$_3$ (1$\bar{1}$02). **d**, Depth-profile of the hydrogen concentration compared for VO$_2$ under microstructure engineering upon identical hydrogenation conditions. **e**, The magnitude of $R_0/R_H$ plotted as the function of hydrogen concentration as compared for VO$_2$ under microstructure engineering.

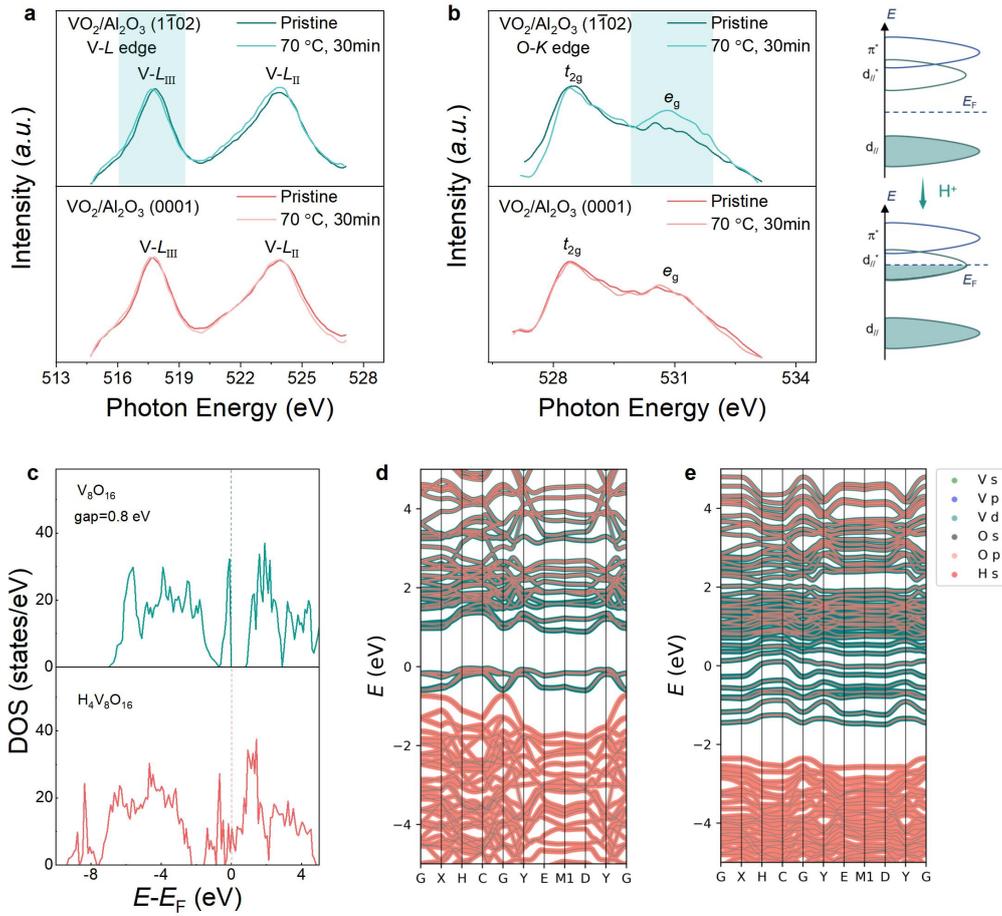

**Figure 4. Band structure for hydrogenated VO$_2$. a-b**, Soft X-ray absorption spectroscopy (sXAS) for the **a**, V-$L$ edge and **b**, O-$K$ edge of VO$_2$/Al$_2$O$_3$ (1$\bar{1}$02) bilayer through hydrogenation, in comparison with the VO$_2$/Al$_2$O$_3$ (0001) heterostructure. The changes in the electronic orbital configuration of VO$_2$ through hydrogenated is schematically shown in the right. **c**, Calculated density of states (DOS) of VO$_2$ film **c**, before and **d**, after hydrogenation. **d-e**, Calculated fat band structure for **d**, pristine VO$_2$, and **e**, H$_4$V$_8$O$_{16}$.